\documentclass[aps,pre,twocolumn,groupedaddress]{revtex4}
\usepackage{epsfig,amsbsy,amsmath,graphicx}
\begin{document}

\title{A simple function for calculating the interaction between a molecule and a graphene sheet}

\author{Xiongce Zhao}
\email{xiongce.zhao@nih.gov}
\affiliation{
Joint Institute for Computational Sciences and 
Center for Nanophase Materials Sciences, Oak Ridge National Laboratory,
Oak Ridge, Tennessee 37831, USA
}
\altaffiliation{Current address: NIDDK, 
National Institutes of Health, Bethesda, MD 20892, USA}


\begin{abstract}

We present a novel potential model for calculating the interaction between
a molecule and a single graphene sheet. The dispersion/repulsion, induction, dipole-quadrupole, 
quadrupole-quadrupole interactions
between a fluid molecule and a graphene sheet are described
by integrated functions that are only dependent on the separation between
the molecule and the graphene along its planar normal. 
The derived potential functions are in excellent
agreement with the computationally demanding atom-explicit summation method. Typical
errors of the integrated potential are less than 2\% in the energy minimum 
compared with the exact atom-explicit summation.
To examine the practical effectiveness of the newly developed functions, 
Monte Carlo simulations were performed to model the adsorption of 
two representative gases in graphene sheets using both the
integrated and atom-explicit potentials.
The integrated potential results in same adsorption isotherms and density profiles
for the adsorbed phase while it only requires negligible computing
time compared with that using the 
atom-explicit method. The newly developed potential functions provide a simple and accurate approach
to calculating the physical interaction between molecules and graphene sheets.

\end{abstract}

\maketitle

\section{Introduction}

Graphene, or individual layers of graphite in which each carbon atom is bonded to three other carbon atoms,
has been the subject of considerable research interest recently \cite{geim07}. 
Being a representative two dimensional crystal, graphene
has many peculiar properties such as low dimensionality, surface homogeneity, 
structure stability, conductivity, and charge transport ability. It is deemed as one of the most
 promising
materials for future applications in various fields 
such as electronics, novel materials, sensors, biodevices etc. \cite{Lee2008,Ponomarenko2008,
wang2008,li2008a,li2008b,
ruoff08,geim07,schedin07,
fasolino07,dikin07,abanin06,novoselov05,zhang05}.

In the investigations of fundamental properties of graphene, it is often of great interest to understand the 
interaction of molecules with graphene on a molecule level. For example, graphene sheets 
have been used as high-sensitivity
sensors \cite{schedin07} or background membranes \cite{Meyer2008} in studying the behavior of 
individual molecules. 
In such applications, 
 interaction between the target molecule and the graphene is one of the key properties that needs to be 
understood in order to
design effective devices. Therefore, a simple and reliable approach to estimating
the graphene-fluid interaction is essential. Furthermore, computer simulation, which is becoming a useful partner of 
experiments in studying fluid-solid interactions, requires realistic and computationally affordable method with sufficient accuracy to calculate
the molecule-solid interactions, such as that between molecules and graphene sheets. 
However, to our knowledge, a simple and accurate function 
describing the interaction between a molecule and a single 
graphene sheet is still not available. In this paper,
we attempt to develop a novel potential to solve this problem.

There are two general approaches to 
calculating the interactions between a molecule and a crystal solid like graphene. 
One can use an atom-explicit potential to describe the interaction between
each atom in the molecule and each atom in crystal, calculating the total potential 
energy by summing up all pairs of atoms. Dispersion, repulsion, electrostatic and induction interactions
 can all be computed in this way. But such an approach is computationally costly, especially 
for large systems involving significant number of solid atoms. Alternatively, one can develop
integrated potential functions that accounts for all the solid atoms in an effective way, making use
of the periodicity and homogeneity in the distribution of atoms in the crystal.
This has been a popular and useful approach for modeling solid-fluid interactions for many
systems, including graphite and metals \cite{steele73}. Integrated potentials are easy to program and
 very efficient computationally.

In this paper the second approach was employed to derive a new set of effective potential force 
fields for molecules
on a graphene sheet. The types of solid-fluid interactions of interest include the dispersion and repulsive
(Lennard-Jones, or LJ), the induction due to the dipole in the
molecule and the polarizability of carbon atoms in graphene, and the interaction between
multipoles in the molecule and the permanent quadrupole in graphene. The  
potential functions derived in this work are extensions of a previous work by us
\cite{zhao05} for modeling interaction between gases and a semi infinite graphitic surface. 
The accuracy of the derived potentials are evaluated by comparing the energies calculated
from the integrated expressions and those from atomistic summation approach. The effectiveness of the potentials 
are demonstrated by using the derived formulas in simulating the adsorption of two representative
gases into slits composed of single graphene sheets.

In Section II, the functions are derived. Section III and IV are examinations of the 
new potentials via comparison with atomistic summation methods. Section V is concluding
remarks.

\section{Potential Development}

We assume that a molecule is located right above the center of a single graphene sheet.
The origin of the system was set as the mass center of the molecule.
If we only consider the pairwise interactions, the total potential energy between the molecule 
and the graphene can be expressed
by a summation
\begin{equation}
\label{eq1}
U=\sum_{i}^\infty u(r_i),
\end{equation}
where $r_i=(x_i,y_i,z)$ denotes the position of a carbon atom in graphene relative to the molecule,
 $u(r_i)$ is the pairwise potential between
the molecule and carbon atom $i$ in graphene, which is only dependent on the separation between 
them, $r_i=(x_i^2+y_i^2+z^2)^{1/2}$. For a graphene sheet of infinite size along its lateral directions ($x$ and $y$),
$U$ only depends on the separation between the molecule and graphene along its surface normal
$z$, and Eq.(\ref{eq1}) becomes
\begin{equation}
\label{eq2}
U(z)=\sum_{i}^\infty u(x_i,y_i,z).
\end{equation}
For example, 
the Lennard-Jones interaction between the molecule and a carbon atom is given as
\begin{equation}
\label{LJ}
u(r)=4\varepsilon_{\mathrm{sf}}\left[\left(\frac{\sigma_{\mathrm{sf}}}{r}\right)^{12}
-\left(\frac{\sigma_{\mathrm{sf}}}{r}\right)^{6} \right],
\end{equation}
where $\varepsilon_{\mathrm{sf}}$ 
and $\sigma_{\mathrm{sf}}$ are the LJ potential parameters for the cross interaction 
between the molecule and a graphene carbon atom. Substituting Eq.(\ref{LJ}) into
Eq.(\ref{eq2}) gives
\begin{equation}
\label{eq3}
U(z)=\sum_i^\infty 4\varepsilon_{\mathrm{sf}}\left[\frac{\sigma_{\mathrm{sf}}^{12}}{(x_i^2+y_i^2+z^2)^6}
-\frac{\sigma_{\mathrm{sf}}^6}{(x_i^2+y_i^2+z^2)^3}\right] ,
\end{equation}
If we assume that the graphene sheet is homogeneous and continuous along its $x$ and $y$ directions, 
$U(z)$ can be approximated by an integral over $x$ and $y$ 
\begin{equation}
\label{eq3a}
U(z)\!=\!4\varepsilon_{\mathrm{sf}}d_\mathrm{s}\!\!\!\int_0^\infty\!\!\!\!\!\int_0^\infty\!\!\!
\left[\frac{\sigma_{\mathrm{sf}}^{12}}{(x^2\!+\!y^2\!+\!z^2)^6}
\!-\!\frac{\sigma_{\mathrm{sf}}^6}{(x^2\!+\!y^2\!+\!z^2)^3}\right]\! dx dy,
\end{equation}
where $d_\mathrm{s}$ is the density of carbon atoms in a unit graphene area $dxdy$. The integration
can be simplified by letting $x^2+y^2=S$, which gives 
 $dxdy=\pi dS$. Then Eq.(\ref{eq3a}) reduces to 
\begin{equation}
\label{eq4}
U(z)=4\pi\varepsilon_{\mathrm{sf}}d_\mathrm{s}\!\!\int_0^\infty\!\!
\left[\frac{\sigma_{\mathrm{sf}}^{12}}{(S+z^2)^6}
-\frac{\sigma_{\mathrm{sf}}^6}{(S+z^2)^3}\right] dS.
\end{equation}
Eq.(\ref{eq4}) leads to an integrated potential for 
the LJ interaction between a molecule and an infinitely large graphene sheet
\begin{equation}
\label{LJsum}
U_{\mathrm{LJ}}(z)=2\pi\varepsilon_{\mathrm{sf}} \sigma_{\mathrm{sf}}^2d_{\mathrm{s}}
\left[\frac{2}{5}\left(\frac{\sigma_{\mathrm{sf}}}{z}\right)^{10}-\left(\frac{\sigma_{\mathrm{sf}}}{z}\right)^4\right].
\end{equation}

The derived expression Eq.(\ref{LJsum}) for the LJ interaction between a molecule and a graphene is
analogous to a potential by Steele \cite{steele73}. The interaction is only a function of the
distance of the molecule from the graphene along its planar normal, $z$, thus the numerical calculation is substantially
simplified. It is noteworthy to point out that unlike the potential for graphite \cite{steele73} the new function for graphene,  
Eq.(\ref{LJsum}), does not contain any empirical term that requires fitted parameters. This is due to the fact that no approximation was involved in the derivation to account for the semi-infinite layers of graphene for a graphite surface. 
 Only assumption here
is that the atoms in graphene sheet are continuous, otherwise the derivation is strictly exact.  

For LJ particles, Eq. (\ref{LJsum}) should be a sufficient approximation for calculating the
molecule-graphene potential. However, for molecules that has partial charges or permanent multipole moments,
additional polar interactions have to be included. For this purpose, similar procedures
are applied to derive integrated forms for the induction and multipolar interactions
between the molecule and a graphene sheet. We note that the above derivation procedure is applicable
 as long as the pairwise potentials are dependent only of
$r$ by an inverse power $r^{-n}$ with $n>1$. 
This condition is satisfied if we use point or angle-averaged dipoles and 
point quadrupoles for the molecule-graphene interaction. For example, the angle-averaged dipole-induced dipole
interaction between a polar molecule and a carbon atom in the graphene is given by \cite{maitland81}
\begin{equation}
\label{ind}
u_{\mu}(r)=-\frac{\mu_{\mathrm{f}}^2\alpha_{\mathrm{s}}}{(4\pi\varepsilon_0)^2r^6},
\end{equation}
where $\mu_\mathrm{f}$ is the permanent dipole moment of the molecule, $\alpha_\mathrm{s}$ is the isotropic
polarizability of a carbon atom in graphene, and $\varepsilon_0$ is the vacuum permittivity. 

The angle-averaged dipole-quadrupole interaction is
given by \cite{maitland81}
\begin{equation}
\label{dq}
u_{\mu\Theta}(r)=-\frac{\mu_{\mathrm{f}}^2\Theta_{\mathrm{s}}^2}{kT(4\pi\varepsilon_0)^2r^8},
\end{equation}
where $\Theta_\mathrm{s}$ is the permanent quadrupole moment on each carbon atom in 
graphene, $k$ is Boltzmann's constant and
$T$ is the absolute temperature. 

Finally, the angle-averaged quadrupole-quadrupole interaction for the molecule and
carbon atom is given by \cite{maitland81}
\begin{equation}
\label{qq}
u_{\Theta\Theta}(r)=-\frac{14\Theta_{\mathrm{f}}^2\Theta_{\mathrm{s}}^2}{5kT(4\pi\varepsilon_0)^2r^{10}},
\end{equation}
where $\Theta_{\mathrm{f}}$ is the permanent quadrupole moment of the molecule.

Substituting Eqs. (\ref{ind})$-$(\ref{qq}) into Eq.(\ref{eq2}) and integrating, 
we obtain the following expressions. The integrated induction potential is 
\begin{equation}
\label{indsum}
U_{\mu}(z)=-\frac{\pi d_{\mathrm{s}}\mu_{\mathrm{f}}^2\alpha_{\mathrm{s}}}{2(4\pi\varepsilon_0)^2}\frac{1}{z^4}.
\end{equation}
The integrated dipole-quadrupole potential is
\begin{equation}
\label{dqsum}
U_{\mu\Theta}(z)=-\frac{\pi d_{\mathrm{s}}\mu_{\mathrm{f}}^2\Theta_{\mathrm{s}}^2}{3kT(4\pi\varepsilon_0)^2}\frac{1}{z^6}.
\end{equation}
The integrated quadrupole-quadrupole potential is
\begin{equation}
\label{qqsum}
U_{\Theta\Theta}(z)=-\frac{7\pi d_{\mathrm{s}}\Theta_{\mathrm{f}}^2\Theta_{\mathrm{s}}^2}
{10kT(4\pi\varepsilon_0)^2}\frac{1}{z^8}.
\end{equation}

Likewise, the potential functions in Eq.(\ref{indsum})$-$(\ref{qqsum}) are 
only dependent on the separation between the multipole and graphene along its planar normal
$z$. By using Eq.(\ref{LJsum}) and Eqs.(\ref{indsum})$-$(\ref{qqsum}), one can calculate
the interaction potential energy between a polar molecule and a graphene sheet 
without using the costly pairwise atomistic summation. 

\section{Comparison with atomistic potentials}

To examine the accuracy of the derived functions, we calculated the potential energies of 
several representative nonpolar and polar molecules
interacting with a single graphene sheet using the integrated functions Eqs.(\ref{LJsum}) and
(\ref{indsum})$-$(\ref{qqsum}), to compare with those from atomistic summation of potentials
Eqs. (\ref{LJ}) and (\ref{ind})$-$(\ref{qq}). The graphene sheet in the atomistic summation 
was modeled as a square about 10 nm in a side, containing 3680 carbon atoms. Trial calculations indicate that such 
a system size contains more than enough carbon atoms to approximate a graphene sheet that is infinite
along its lateral directions. A single molecule was placed over the center of the graphene sheet and its 
interaction potential energies were computed by summing up interactions from each carbon atom. 

\begin{table}
\caption{
The Lennard-Jones and multipole parameters for the example molecules studied.
}
\label{tab1}
\begin{ruledtabular}
\begin{tabular}{lcccc}
	&$\varepsilon_{\mathrm{f}}$ (kcal/mol)	&$\sigma_{\mathrm{f}}$ (\AA)	
&$\mu_{\mathrm{f}}$ (Debye)	&$\Theta_{\mathrm{f}}$ (10$^{-20}$C \AA$^2$) \\

\hline
CH$_4$	&0.2378	&3.527	&$-$	&$-$\\
H$_2$O	&0.1554	&3.152	&2.351	&$-$\\
Cl$_2$	&0.7101	&4.115	&$-$	&10.79\\
\end{tabular}
\end{ruledtabular}
\end{table}

The LJ and multipole potential parameters for the example molecules chosen are shown in Table \ref{tab1}. 
The nonpolar molecule studied is methane, modeled by a single LJ site \cite{jiang93}. 
The representative polar molecules 
selected are water and chlorine. Water is modeled by the SPC/E potential \cite{Berendsen87}, 
with a strong dipole moment of 2.351 Debye.
Please note that the LJ parameters of water in Table \ref{tab1} are for the O atom. 
 The chlorine molecule is modeled by a single LJ site plus a quadrupole 
moment of 10.79$\times10^{-20}$ C \AA$^2$ \cite{rowley}. 
The purpose of the calculation was to study the accuracy of the integrated potential 
compared with atomistic summation. 
Therefore, the potential models for the molecules were selected arbitrarily. The values of potential 
parameters for carbon atoms
in graphene are $\sigma_{\mathrm{s}}$=3.40 \AA, $\varepsilon_{\mathrm{s}}$=0.05569 kcal/mol, 
$d_{\mathrm{s}}$=0.382 \AA$^{-2}$ \cite{zhao05}, $\alpha_{\mathrm{c}}$=1.76 \AA$^3$ \cite{bates77}, 
$\Theta_{\mathrm{s}}$=$-3.03\times$10$^{-20}$ C \AA$^2$ \cite{whitehouse93}. The cross interaction 
parameters $\varepsilon_{\mathrm{sf}}$ and 
$\sigma_{\mathrm{sf}}$ were calculated using the Lorentz-Bertholet rules. 

\begin{figure}
\includegraphics[width=80mm]{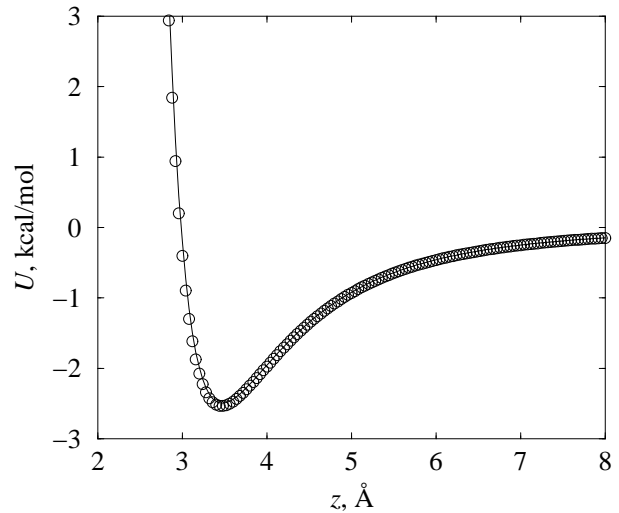}\\%
\caption{
Interaction of methane with a single graphene sheet calculated using the atom-explicit
pairwise summation of Eq.(\ref{LJ}) (line) and the integrated function Eq.(\ref{LJsum}) 
(symbols).
}
\label{methane}
\end{figure}

\begin{figure}
\includegraphics[width=80mm]{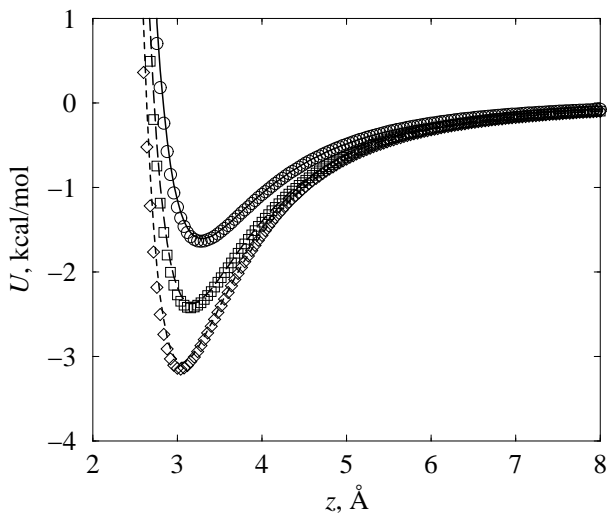}\\%
\caption{
Interaction of a water molecule with a single graphene sheet calculated using the atomistic
summation (lines) and integrated functions (symbols).
Atomistic summations: LJ interaction Eq.(\ref{LJ}) (solid line), LJ plus induction 
interactions
Eq.(\ref{LJ})$+$Eq.(\ref{ind}) (long dashed line), and LJ plus induction plus
dipole-quadrupole interactions Eq.(\ref{LJ})$+$Eq.(\ref{ind})$+$Eq.(\ref{dq}) (dashed line).
Integrated functions: LJ interaction Eq.(\ref{LJsum}) (circle), LJ plus induction interactions
Eq.(\ref{LJsum})$+$Eq.(\ref{indsum}) (square), and LJ plus induction plus
dipole-quadrupole interactions Eq.(\ref{LJsum})$+$Eq.(\ref{indsum})$+$Eq.(\ref{dqsum}) (diamond).
}
\label{water}
\end{figure}

\begin{figure}
\includegraphics[width=80mm]{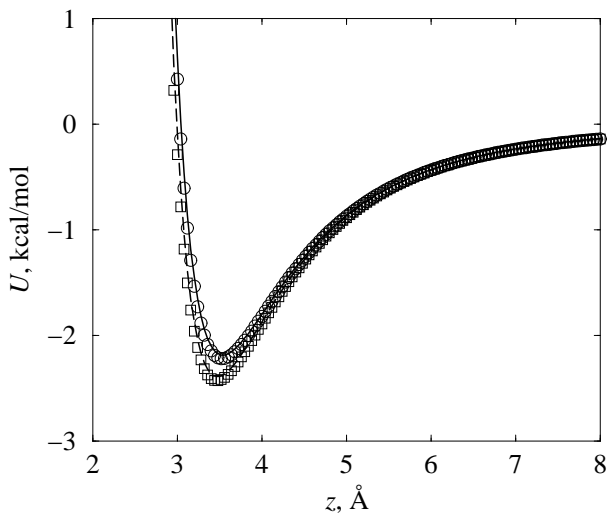}\\%
\caption{
Interaction a chlorine molecule with a single graphene sheet calculated using the atomistic
summation (lines) and integrated functions (symbols).
Atomistic summations: LJ interaction Eq.(\ref{LJ}) (solid line), LJ plus quadrupole-quadrupole interactions
Eq.(\ref{LJ})$+$Eq.\ref{qq} (long dashed line).
Integrated functions: LJ interaction Eq.(\ref{LJsum}) (circle), LJ plus quadrupole-quadrupole interactions
Eq.(\ref{LJsum})$+$Eq.(\ref{qqsum}) (square).
}
\label{chlorine}
\end{figure}

In Figs. \ref{methane} to \ref{chlorine} we compare the potential energies calculated from the integrated
potentials and the atom-explicit potentials at 300 K, for CH$_4$, H$_2$O, and Cl$_2$. It is seen that
the integrated expressions are in excellent agreement with the atomistic summation results. If we take
the atomistic summation results as the standard, the typical errors in the integrated potential are 
less than 2\% at the energy minimum. The integrated functions always give a slightly more attractive
potential compared to the atom-explicit methods, possibly due to the fact that we treat the graphene carbon atoms as
continuum in the integrated potential. 

The integrated potentials are most accurate when the graphene sheet is infinitely large. However, we also
carried out series of calculations using graphene sheets of various sizes, to test the 
applicability of the derived functions to the finite-size graphene. 
We found that the integrated potentials are in excellent agreement with the atomistic potentials
as long as the square-shaped graphene sheet contains 
more than $\sim$500 carbon atoms (or a graphene sheet of $\sim$38$\times$38 \AA$^2$). 
This suggests that the derived potentials can also be readily 
used in estimating the interaction of molecules with finite
graphene sheets that are nanometer-scale in size. 

For the different types of interactions between a molecule and graphene, it is interesting to see that,
for example from Fig. \ref{water}, the induction and dipole-quadrupole interactions for strongly polar molecules
such as water are non-negligible compared with the LJ energy. On the other hand, based on the Cl$_2$
example, the quadrupole-quadrupole interaction is usually not significant, even for 
molecules with a strong quadrupole moment like chlorine. 

\section{Applications in adsorption simulation}

One of important applications of the potential functions developed in this work is molecular simulation 
of adsorption of molecules in pores composed of single graphene sheets. 
In turn, such simulations serve as a verification of the accuracy and effectiveness of the integrated functions
in practical applications.
For this purpose, we chose to simulate the adsorption of
two typical nonpolar and polar gases, methane and water, confined in single graphene sheets. 
The simulations were performed
using the grand canonical Monte Carlo (GCMC) method, the detailed description 
of which can be found in literature \cite{allen}.

The potentials parameters describing the interactions of water and methane with graphene are given in
Table \ref{tab1}. The methane is modeled as a simple one-site LJ particle, with a cutoff
of 9 \AA. The LJ interaction between water molecules is also modeled by a cutoff of 9 \AA, without
long-range correction applied, as suggested by the original literature \cite{Berendsen87}.
The electrostatic interaction between water molecules are modeled by 
the partial charges distributed on the three charge sites on the water model, without long range correction.
The interaction between the molecules and graphenes is calculated by either the atomistic
summation approach or integrated functions for comparison. 

The GCMC cell for adsorption simulations is a rectangular box. Two graphene sheets are placed in the cell
with their planes parallel to the $x-y$ plane of the box. The distance between the two graphene layers
along the $z$ direction is 15 \AA. Again, this separation is chosen arbitrarily. 
Periodic boundary conditions
are applied in all three directions. Therefore, 
the height of simulation box in the $z$ direction is
30 \AA. The typical box sizes in the $x$ and $y$ directions are set as 
32 and 34 \AA\ respectively, leading to a cell volume of about 33 nm$^3$. 
The types of move attempted during a
GCMC simulation were selected randomly with probability of 0.40, 0.40, 0.10,
and 0.10 for displacements, rotations, creations, and deletions of a fluid 
molecule respectively. For simulation of the spherical methane, the rotation moves were merged
to displacement moves.  Each simulation included equilibration of 2$\times10^6$
MC moves and production of 2$\times10^6$ MC moves. 

The simulation of methane adsorption in graphene sheets were performed at 77 K \cite{jiang93}.
Previous theoretical studies predict that water only wets graphitic carbon surfaces when temperature is
above $\sim$510 K \cite{gatica04,zhao07}. Therefore,
we chose to perform simulations of water/graphene at 550 K.
In GCMC simulations the reduced chemical potential was varied to obtain isotherms and density profiles
of adsorbed methane or water, using either atomistic or integrated 
 potential for comparison. 

\begin{figure}
\includegraphics[width=80mm]{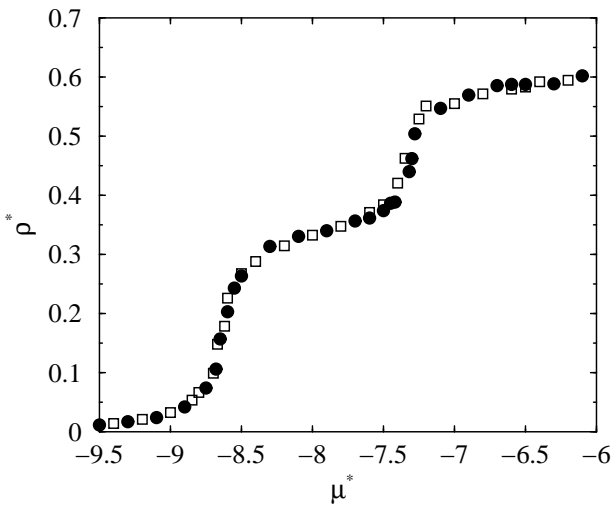}\\%
\caption{
Adsorption isotherm of methane in singlet graphene sheets at 77 K.
Circles were calculated using the atomistic summation potential and squares were calculated using the
integrated potential. Adsorbed amount ($\rho^*$) and chemical potential ($\mu^*$) are in
reduced unit.
}
\label{methaneI}
\end{figure}

\begin{figure}
\includegraphics[width=80mm]{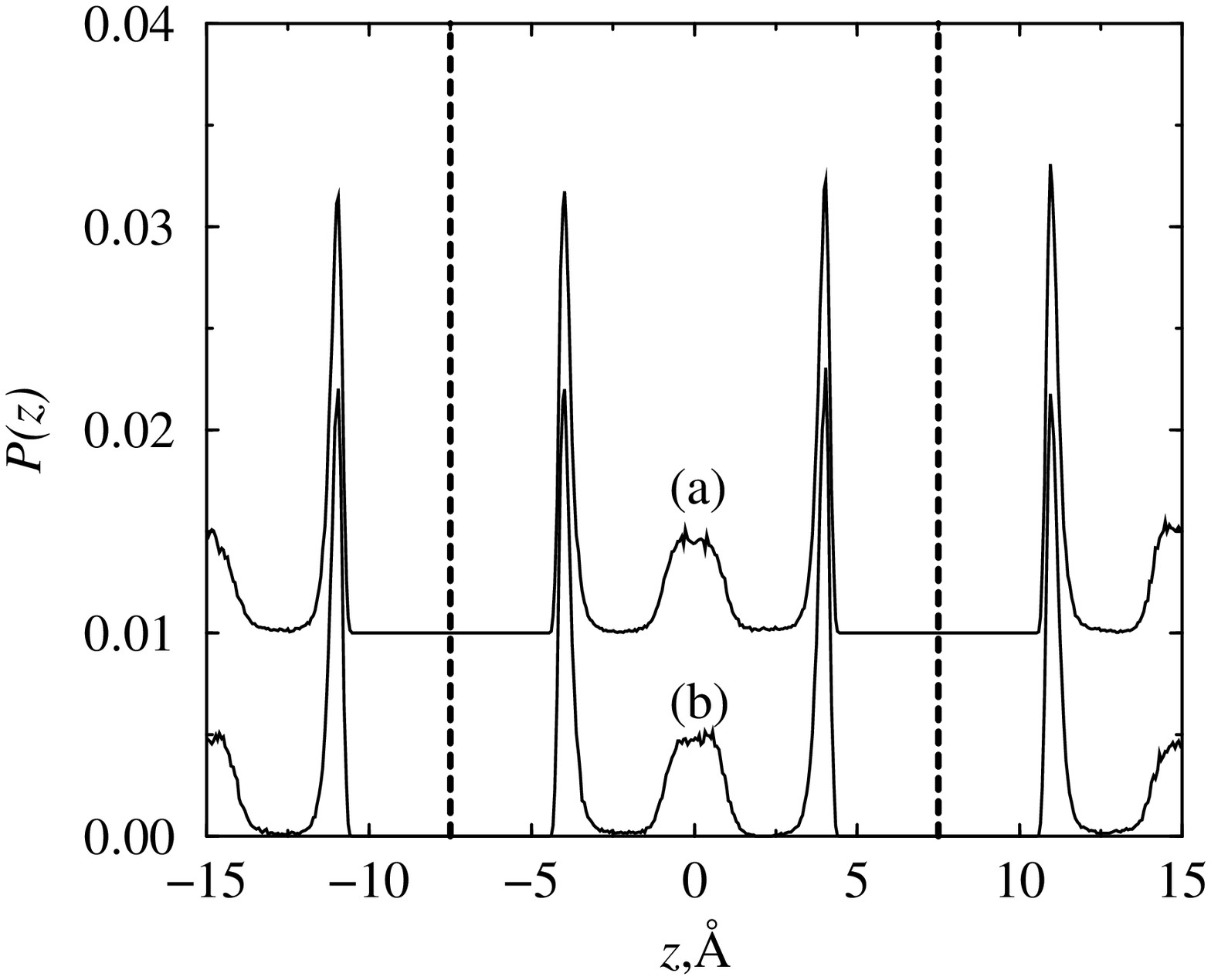}\\%
\caption{
Density profiles of methane adsorption in graphene sheets at 77 K and reduced chemical potential of
$-6.5$. (a) was calculated using the atomistic summation potential and (b) was calculated using the
integrated potential. Two curves are almost identical so (a) was shifted by 0.01 {in} $P(z)$ for clarity.
The dashed lines at $z=\pm7.5$ \AA\ represent the two graphene sheets in the system.
}
\label{methaneP}
\end{figure}

\begin{figure}
\includegraphics[width=80mm]{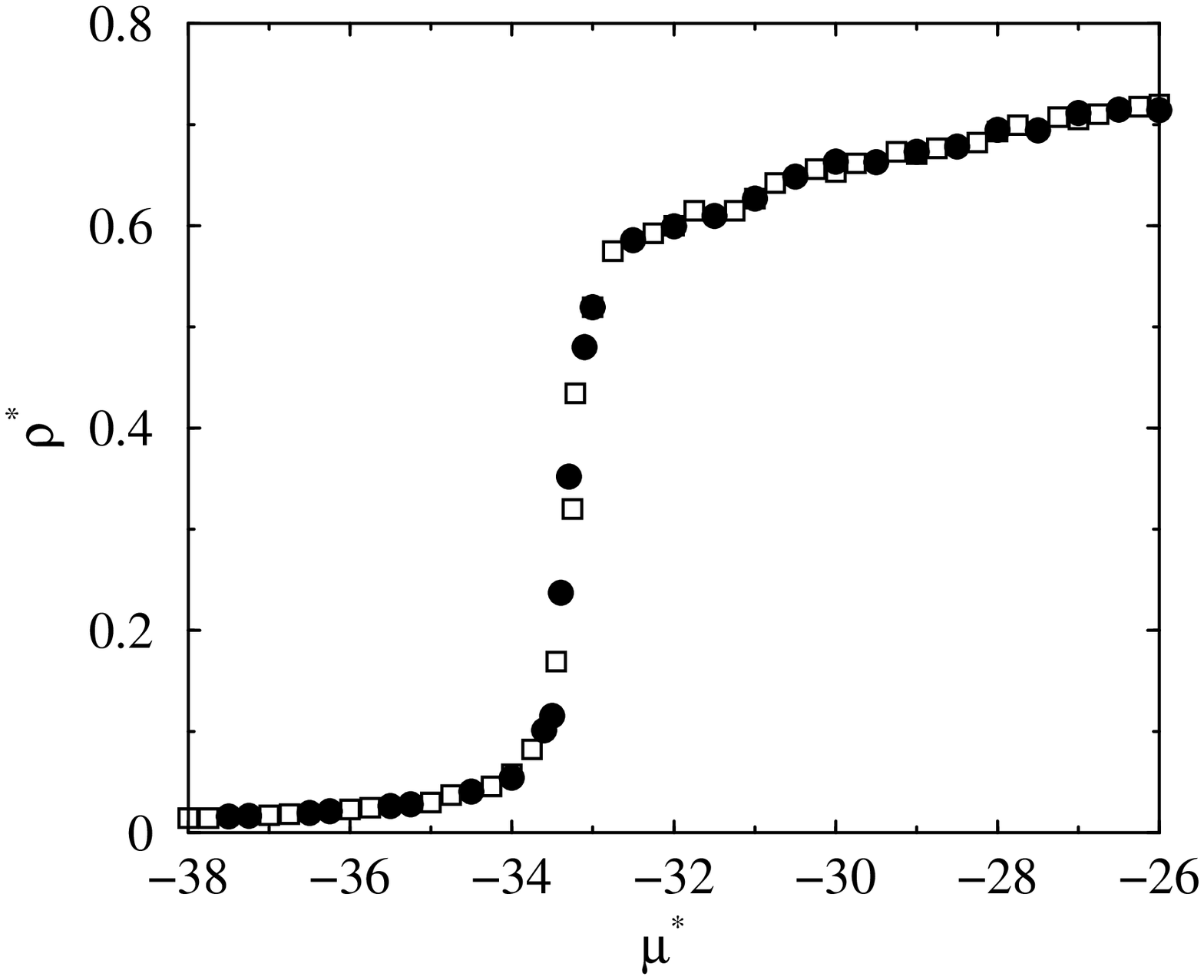}\\%
\caption{
Adsorption isotherm of water in graphene sheets at 550 K.
Circles were calculated using the atomistic summation potential and squares were calculated using the
integrated potential. Adsorbed amount ($\rho^*$) and chemical potential ($\mu^*$) are in
reduced unit. 
}
\label{waterI}
\end{figure}

\begin{figure}
\includegraphics[width=80mm]{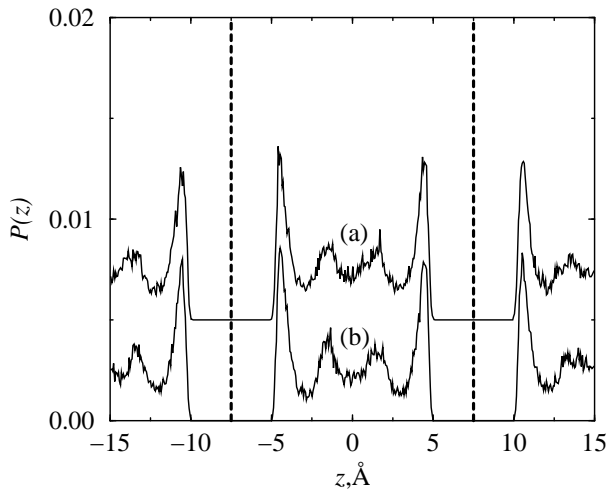}\\%
\caption{
Density profiles of water adsorption in graphene sheets at 550 K and reduced chemical potential of
$-26$. (a) was calculated using the atomistic summation potential and (b) was calculated using the
integrated potential. Two curves are almost identical so (a) was shifted by 0.005 {in} $P(z)$ for clarity.
The dashed lines at $z=\pm7.5$ \AA\ represent the two graphene sheets in the system.
}
\label{waterP}
\end{figure}

Calculated adsorption isotherms for methane in graphene sheets at 77 K are presented in Fig. \ref{methaneI}. 
The circles are simulation
results computed from the atomistic summation of Eq.(\ref{LJ}), while
the squares are from simulations using the integrated potential Eq.(\ref{LJsum}).
It can be seen that, within statistical uncertainty, 
the isotherms from these two different potentials are in excellent agreement.
 The first plateau in the isotherm corresponds to the first layers
of methane adsorbed on graphene sheets, and the second plateau is from
the methane adsorbed onto first layers as chemical potential increases. This is confirmed by  
density profile distribution of the adsorbed phase shown in Fig. \ref{methaneP}. 
The transition from nonadsorption to the first layer occurs at about $\mu^*=-8.6$, 
and the transition from the first to the second layer occurs at
about $\mu^*=-7.3$.

It is known that the isotherm shapes, especially the 
layering transition of adsorbed phase calculated from simulations, are very sensitive 
to the solid-fluid interactions \cite{zhao02}. A small deviation in the solid-fluid 
potential can result in significant shift in the layer transition location in isotherms. 
The excellent agreement in the adsorption layering transition for the two potentials indicates that the 
integrated functions are robust alternatives to the atomistic models. 

One example of density profiles of methane adsorbed
in graphene layers is given in Fig. \ref{methaneP}. Sharp peaks are observed
at $z=\pm4$, $\pm11$ \AA, which correspond to the first adsorbed methane layers on the graphene sheets.
Additional broad peaks are observed at $z=0$, $\pm15$ \AA, representing the adsorbed second layers between
the first layers. 
Again, it is seen that the density distribution of adsorbed
methane in graphenes calculated using the atomistic and integrated potentials are in excellent agreement.
 Note that the density profiles calculated from the two potentials are almost identical to each other, 
so that we have to present them by shifting
one curve along the vertical axis by a constant value to have a clear view. 

The results for water adsorption in singlet graphenes at 550 K
are shown in Figs. \ref{waterI} and \ref{waterP}. It is seen from the water adsorption isotherm
that a continuous wetting transition occurs at about $\mu^*=-33.4$. The density profile peaks corresponding to the
first adsorbed layers on the graphene sheets locate
at $z=\pm4.4$ and $\pm10.6$ \AA. Slightly different from that of methane,
the adsorbed phase of water forms two second layers between the first layers, at $z=\pm1.5$ and $\pm13.5$ \AA, 
due to the relatively
smaller size of water molecule compared to methane. The isotherms and density profiles
of water in graphene sheets calculated from the integrated and atomistic potentials are also 
in excellent agreement. This indicates that the integrated potentials for multipolar interactions
are also excellent approximations to the atomistic approach. 

We monitored the computational time required for the simulations using two different approaches. It is 
found that typically the computational time for the integrated potential 
 is $<1\%$ of that required for the atomistic summations.

\section{Conclusions}

A set of effective potentials were derived for calculating the Lennard-Jones,
dipole-induced dipole, dipole-quadrupole, and quadrupole-quadrupole interactions
between fluid molecules and a graphene sheet. The integrated
potential functions depend only on the separation between a molecule and the graphene surface.
They are mathematically simple and easy to use in either estimating the 
interaction between a single molecule and a graphene sheet
 or in large-scale molecular simulations. 
The potential energies calculated from the integrated potentials are in excellent agreement 
with the results calculated from direct atomistic summations, while the integrated
ones are computationally negligible compared with their atomistic counterparts.
Adsorption simulations of two representative gases in singlet graphene sheets
were performed to further test the derived potential functions. The adsorption isotherms and density profiles
calculated from the derived potentials and atomistic models are in excellent agreement, indicating
that the derived potential can predict both the equilibrium and the structural properties
of adsorbed phase with excellent accuracy. The potentials
developed in this work provide a simple, accurate, and robust method for calculating
the physical interaction of molecules and single graphene sheets.

Finally, we note that the self-consistency in the polarization of carbon atoms by a polar fluid
molecule was neglected in the derivation. However, we believe that the effect is relatively small compared with
the dispersion, induction, and dipole-quadrupole interactions. Also, by using an integrated
potential, one assumes the graphene surface is smooth and the impact of the surface corrugation
on the molecule-solid interaction is ignored.

\begin{acknowledgments}
The author thanks Peter T. Cummings
for many insightful discussions. This research was
conducted at the Center for Nanophase Materials Sciences, which is sponsored
at Oak Ridge National Laboratory by the Division of Scientific User
Facilities, U.S. Department of Energy.
This research used resources of the National Energy Research Scientific
Computing Center, which is supported by the Office of Science of
the U.S. Department of Energy under Contract No. DE-AC02-05CH11231.

\end{acknowledgments}

\providecommand{\url}[1]{\texttt{#1}}
\providecommand{\refin}[1]{\\ \textbf{Referenced in:} #1}


\end{document}